\documentclass[aps,floatfix,prx,twoside,twocolumn,10pt,showpacs, citeautoscript, superscriptaddress,]{revtex4-1}

\usepackage{amsmath}
\usepackage{amssymb}
\usepackage{booktabs}
\usepackage{comment}
\usepackage{glossaries}
\usepackage{graphicx}
\usepackage[version=4]{mhchem}
\usepackage{siunitx}
\usepackage{tabularx}
\usepackage[colorlinks=true]{hyperref}
\hypersetup{
    pdffitwindow=false,
    pdfstartview={FitH},
    pdfnewwindow=true,
    colorlinks=true,
    linkcolor=blue,
    citecolor=blue,
    filecolor=blue,
    urlcolor=blue,
}

\hypersetup{pdfauthor={Someone}}
\hypersetup{pdftitle={Some title}}

\setacronymstyle{long-short}
\newacronym{acf}{ACF}{autocorrelation function}
\newacronym{dft}{DFT}{density functional theory}
\newacronym{gf}{GF}{generating function}
\newacronym{gpu}{GPU}{graphical processing unit}
\newacronym{md}{MD}{molecular dynamics}
\newacronym{mlp}{MLP}{machine-learning potential}
\newacronym{nep}{NEP}{neuroevolution potential}
\newacronym{pes}{PES}{potential energy surface}
\newacronym{zpl}{ZPL}{zero-phonon line}

\DeclareSIUnit\angstrom{\text{\AA}}

\newcommand{\addChalmers}{Chalmers University of Technology, Department of Physics, 41296 Gothenburg, Sweden}

\begin{document}
\title{Direct, indirect, and self-trapped excitons in \ce{Cs2AgBiBr6}}

\author{Mehmet Baskurt}
\affiliation{\addChalmers}
\author{Paul Erhart}
\author{Julia Wiktor}
\email{julia.wiktor@chalmers.se}

\affiliation{\addChalmers}

\begin{abstract}
\ce{Cs2AgBiBr6} exhibits promising photovoltaic and light-emitting properties, making it a candidate for next-generation solar cells and LED technologies. Additionally, it serves as a model system within the family of halide double perovskites, offering insights into the broader class of materials. 
Here, we study various possible excited states of this material to understand its absorption and emission properties.
We use Time-Dependent Density Functional Theory (TD-DFT) coupled with non-empirical hybrid functionals, specifically PBE0($\alpha$) and dielectric-dependent hybrids (DDH) to explore direct, indirect, and self-trapped excitons in this material.
Based on comparison with experiment, we show that these methods can give excellent prediction of the absorption spectrum and that the fundamental band gap has been underestimated in previous computational studies.
We connect the experimental photoluminescence signals at 1.9-2.0 eV to the emission from self-trapped excitons and electron polarons.
Finally, we reveal a complex landscape with energetically competing direct, indirect, and self-trapped excitons in the material.
\end{abstract}

\maketitle
Halide perovskites (HPs), materials with the general $ABX_3$ formula, have attracted significant interest for their advanced optoelectronic applications, including solar panels and lighting.
A subclass within the HP family, halide double perovskites, replaces the single $B$ metal cation with a combination of two different elements, yielding the $A_2BB'X_6$ formula.
Also known as elpasolites, these materials hold potential for applications in photovoltaics, X-ray detection, and white light emission.
Among halide double perovskites, \ce{Cs2AgBiBr6} stands out as one of the most studied.
It has gained interest as a viable alternative to lead-based perovskites due to its high stability, non-toxicity, outstanding optoelectronic properties, and multifunctionality.
Additionally, it has emerged as a benchmark case, both for experimental and computational studies.

One of interesting properties of halide double perovskites is that they often exhibit significant light emission.
In \ce{Cs2AgBiBr6}, a photoluminescence (PL) peak has been measured at 1.9-2.0~eV\cite{zelewski2019revealing, schade2018structural, wright2021ultrafast}.
This emission has been assigned to the indirect band gap~\cite{slavney2016bismuth, schade2018structural, palummo2020optical} or sub-gap states~\cite{hoye2018fundamental} such as color-centers~\cite{zelewski2019revealing, wright2021ultrafast}.
It is therefore useful to computationally assess the likelihood of different sources of emission.
Since the possible interpretations involve both free (in particular indirect) excitons and self-trapping, such comparison requires a method that can model energetics of delocalized and localized excitations on the same footing.

The study of excited states requires methods extending beyond the density functional theory (DFT), such as the Bethe-Salpeter equation (BSE)\cite{albrecht1998ab, rohlfing1998excitonic, benedict1998optical, onida2002electronic, sun2023toward}. 
However, the high computational cost of BSE makes its use impractical in the case of self-trapped excitons (STEs), which in the most commonly used approach require the use of supercells.
One notable exception is a study by Ismail \textit{et al.} \cite{ismail2005self} on the STE in SiO$_2$.
We note that recent developments by Dai \textit{et al.}\cite{dai2024excitonic1, dai2024theory} made it possible to model STEs in unit cells.
However, this technique has not been yet widely applied and in the present study we want to focus on a cubic perovskite phase, which has been shown to be impossible to describe completely using a small symmetric structure~\cite{wiktor2017predictive, cs2agbibr6_polymorphous, wiktor2023quant}. Alternatively, recent advancements have allowed for more computationally efficient alternatives without compromising the accuracy inherent in BSE.
Most notably, the combination of time-dependent density functional theory (TD-DFT) with non-empirical hybrid functionals has been shown to achieve the accuracy of BSE at a fraction of the cost \cite{sun2020low, tal2020accurate}.
While this method has been so far used for pristine materials and free excitons, its predictive power and relatively high computational efficiency make it a promising method for studying STEs as well.
In a recent study by Jin~\textit{et al.}~\cite{jin2024self} it has been shown that STEs can be efficiently studied using the TD-DFT method.
At the same time, they showed that the constrained-occupation DFT method, also called $\Delta$SCF, leads to very similar geometries using TD-DFT forces to relax the structure of the STE.

In the present study, we combine TD-DFT with two types of non-empirical hybrid functionals.
One is based on the PBE0($\alpha$) functional where the fraction of exact exchange $\alpha$ is set to a value which satisfies the generalized Koopmans' condition~\cite{koopmans_1, koopmans_2, baskurt2023charge}.
Another one belongs to the class of dielectric-dependent hybrid functionals (DDH)~\cite{chen2018nonempirical, cui2018doubly, liu2019assessing}, in which the exchange potential follows the inverse of the dielectric function.
From these methods, we extract the transitions of the direct and indirect free excitons as well as of the self-trapped exciton in the singlet and triplet state.
We show that different excited states can be close in energy and by performing careful convergence studies we obtain the emission energy from the singlet state of the STE close to the experimental PL signal.

\begin{figure*}[t]
    \centering
    \includegraphics[width=\linewidth]{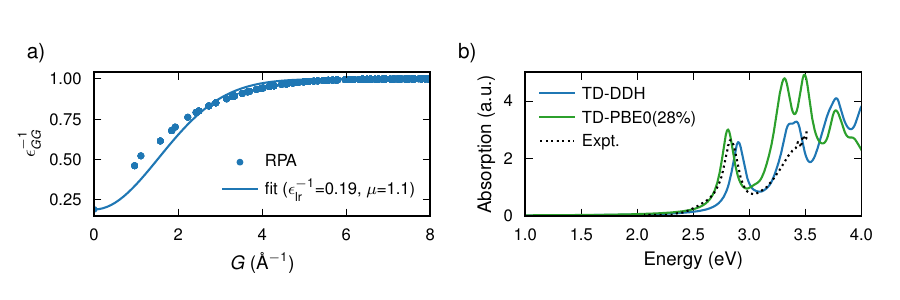}
    \caption{
    (a) Inverse dielectric function versus wave vector at the $\Gamma$ point calculated within the RPA and fitted with Eq.~\ref{eq1}.
    (b) Absorption spectra calculated using TD-DDH and TD-PBE0(28\%) compared with the experimental results from Ref.~\citenum{longo2020understanding}. Computed absorption spectra include a convolution with Lorentzians with a width of 0.07~eV and were renormalized to match the intensity of the first experimental peak.
    }
    \label{fig:fig1}
\end{figure*}
Calculations are performed using two simulation cells. Parametrization of the DDH functional, most convergence tests, and the comparison with the experimental absorption spectra are based on the unit cell of \ce{Cs2AgBiBr6} (spacegroup F$m\bar{3}m$) using the experimental lattice parameter of 11.27~\AA{}~\cite{cs2agbibr6_latticeconstant}.
Calculations for the free excitons and the STE are done in a supercell containing 320 atoms, based on the polymorphous cubic structure \cite{cs2agbibr6_polymorphous} from Ref.~\citenum{baskurt2023charge}, where we used the $\Delta$SCF method for the excited state. In that work, we studied charge localization in \ce{Cs2AgBiBr6} and showed that small electron polarons and STEs are stable in the material, while the localization of small hole polarons is less favorable. We note that these results align well with the later findings by Lafuente-Bartolome \textit{et al.}\cite{lafuente2024topological}.
From Ref.~\citenum{baskurt2023charge}, we adopt the $\alpha$ parameter in PBE0($\alpha$) of 0.28, which has been shown to satisfy the Koopmans' condition for the Br vacancy.
All simulations presented in the main text include spin-orbit coupling (SOC), while some of the convergence tests shown in the Supporting Information (SI) are done without this contribution.
Additional computational details and convergence tests can be found in the SI.

We first calculate the dielectric function $\epsilon$ of \ce{Cs2AgBiBr6} within the random-phase approximation (RPA).
The calculation is carried out using an energy cutoff of 400~eV and a $k$-point mesh of 6$\times$6$\times$6.
As shown in the SI, these parameters lead to a very well converged dielectric function.
The result is given in Fig.~\ref{fig:fig1}a in which we also include the fit following:
\begin{equation}
\epsilon^{-1}(G) = 1 - (1 - \epsilon_{\rm lr}^{-1})e^{-G^2/4\mu^2},
\label{eq1}
\end{equation}
where $\epsilon_{\rm lr}^{-1}$ corresponds to the long-range exchange fraction at $G=0$, $G$ is the wave vector, and $\mu$ is the range-separation parameter.
We obtain $\epsilon_{\rm lr}^{-1}=0.19$ and $\mu=1.1$.

To verify the validity of our hybrid functionals, we compare the absorption spectra calculated via TD-DFT with experimental results from Ref.~\citenum{wright2021ultrafast} (see Fig.~\ref{fig:fig1}b). The absorption $\alpha_{\text{abs}}$ is calculated based on the real and imaginary parts of the dielectric function $\epsilon$ as:
\begin{equation}
    \alpha_{\text{abs}}(\omega) = \sqrt{2} \frac{\omega}{c} \sqrt{\left| \epsilon(\omega) \right| - \epsilon_R(\omega)},
\end{equation}
where $c$ is the speed of light and $\omega$ is the angular frequency~\cite{sadigh2011first}.
Convergence tests corresponding to the TD-DFT calculations are given in the SI.
We observe a good agreement with the experimental results using both hybrid functional, with TD-PBE0(28\%) reproducing the position of the first absorption peak almost exactly and TD-DDH slightly overestimating it, while giving a better agreement for the valley and the second peak.
We attribute the overestimation of the second peak within TD-PBE0(28\%) to its incorrect asymptotic behavior.
Using the nonempirical hybrid functionals we obtain a much better agreement with experimental data than previous $G_0W_0$ calculations~\cite{biega2021chemically}, where the position of the first peak was underestimated by about 0.6~eV.
This is due to the fact that while non-empirical hybrid functionals have been shown to give high accuracy in describing band gaps of halide perovskites~\cite{koopmans_2, wang2022accurate}, the one-shot $G_0W_0$  used in Ref.~\citenum{biega2021chemically} can significantly underestimate that property~\cite{filip2014g, wiktor2017predictive, leppert2019towards}. The authors of Ref.~\citenum{biega2021chemically} noted that this underestimation is primarily due to the lack of self-consistency and the starting point dependence of $G_0W_0$ calculations, which is a well-known issue. Although other calculations in Ref.~\citenum{palummo2020optical} considered partial self-consistency by updating energies in $G$ and $W$ (the $evGW$ scheme), this approach still did not lead to the full increase of the band gap that would reproduce experimental results.
In Table~\ref{table1} we include the fundamental direct and indirect gaps calculated with the DDH and PBE0($\alpha$) methods and compare them with previous computationally reported values (the band structure is given in the SI).
First, we note that the two hybrid functionals constructed using different physical considerations lead to almost the same indirect and direct band gaps. Our indirect band gap is also close to the value recently calculated by Wang \textit{et al.} using the double screened hybrid (DSH) functional~\cite{wang2024high}. Second, even when similar methods are employed, such as $G_0W_0$ on top of LDA or PBE, band gaps can differ by as much as 0.35 eV. This discrepancy can be attributed to differences in the underlying functional, codes used, the choice between PAW and norm-conserving pseudopotentials, and the selection of valence states.
Third, we observe that the values present here, about 2.7 and 3.5~eV for the indirect and direct transition, respectively, are higher than most previous estimates. However, considering the excellent agreement between absorption spectra calculated here and the experiment, we argue that the presented fundamental gaps are more reliable predictions than the previously reported values. 

\begin{table}[b]
\centering
\footnotesize
\caption{
    Fundamental indirect and direct band gaps from the PBE0($\alpha$) and DDH functionals compared with previously reported values.
}
\label{table1}
\begin{tabular}{l r r}
&Indirect gap (eV)&Direct gap (eV)\\
\hline
PBE0(28\%)&2.66&3.52\\
DDH&2.66&3.50\\
DSH Ref.~\citenum{wang2024high} &2.50&\\
$G_0W_0$@LDA Ref.~\citenum{filip2016band}& 1.83&2.51\\
$G_0W_0$@PBE Ref.~\citenum{leppert2019towards}& 2.01&\\
$G_0W_0$@LDA Ref.~\citenum{biega2021chemically}&1.66&2.41\\
$G_0W_0$@HSE Ref.~\citenum{leppert2019towards}& 2.59&\\
$GW_0$@HSE Ref.~\citenum{leppert2019towards}& 2.82&\\
$evGW$ Ref.~\citenum{palummo2020optical}&2.1&2.7\\
\end{tabular}
\end{table}

\begin{figure}[h]
    \centering
    \includegraphics[width=0.9\linewidth]{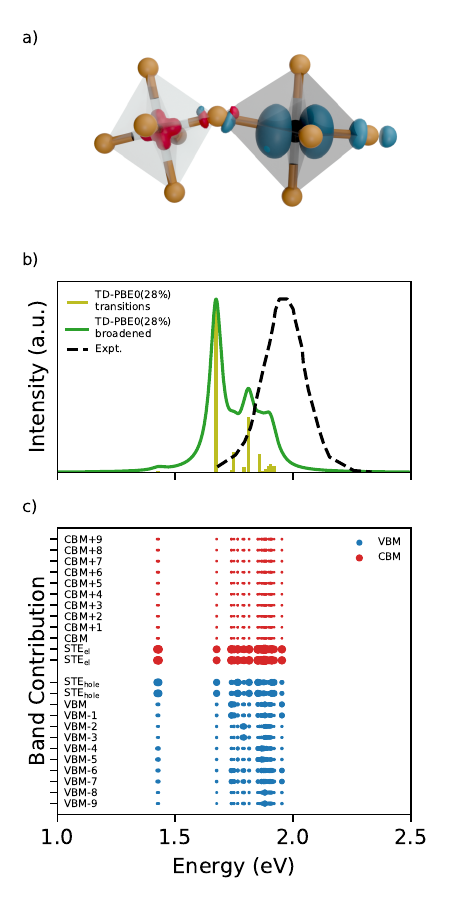}
    \caption{
    (a) Isodensities of the hole (in red) and electron (in blue), localized within the AgBr$_6$ and BiBr$_6$ octahedron, respectively. The rest of the structure was removed for clarity.
    (b) Emission from the STE compared with the experimental PL spectrum from Ref.~\citenum{wright2021ultrafast}.
    The broadened spectrum was generated by convolution with Lorentzians with a width of 0.03~eV. All spectra were rescaled to have the same maximum.
    (c) Band contributions to each of the transitions marked in the upper panel.
    For each initial (final) state, the contributions of different final (initial) states, as well as all $k$-points, are added up.
    }
    \label{fig:fig2}
\end{figure}

The previous calculations were performed for the perfect cubic structure of \ce{Cs2AgBiBr6}.
It has been shown that such a symmetric model is not a good representation of the locally and dynamically disordered halide perovskites~\cite{wiktor2017predictive, cs2agbibr6_polymorphous, wiktor2023quant}. We note that Ref.~\citenum{cohen2022diverging} has shown that unlike lead-based perovskites, \ce{Cs2AgBiBr6} has well-defined normal modes up to room temperature, which means that the symmetric average cubic structure is enough to describe the electronic structure of the material even at finite temperatures. We demonstrate this by comparing the imaginary part of the dielectric function calculated within the pristine F$m\bar{3}m$ unit cell and the polymorphous supercell in the SI. However, introducing local distortions within the perfect cubic supercell would either lead to its relaxation to a form of a polymorphous cell or to finding a local minimum where the optimal charge localization is not possible~\cite{baskurt2023charge}.
Therefore, we now turn to the study of the excited states of \ce{Cs2AgBiBr6} based on a more realistic polymorphous model.
We adopt one polymorphous structure for the pristine \ce{Cs2AgBiBr6} and one for the STE from Ref.~\citenum{baskurt2023charge}. We note that we have also tested how the properties of the STE change when the low-temparature tetragonal structure (I4/m) is considered and found that similar transitions can be found in that model (see SI for the test).
These structures were generated using the \textsc{cp2k} code~\cite{cp2k_1,cp2k_2} and here, for consistency, we further relax the atomic position within \textsc{vasp}.
In the following, we focus on the PBE0(28\%) functional, as it gives a slightly better agreement with experiment for the lower part of the absorption spectrum.
The STE is relaxed within its lowest triplet state and we assume that the atomic positions do not change significantly for the singlet state. We show the isodensities of the localized hole and electron in Figure~\ref{fig:fig2}a).
We then perform TD-DFT calculation using the two hybrid functionals on top of these geometries.
While the relaxation is performed using the $\Gamma$ point only, for the TD-DFT calculations we use a special grid with four $k$-points.
As shown in the SI, this grid yields lower parts of the absorption spectra that agree well with those from the 2$\times$2$\times$2 grid. The differences in the positions of the two lowest peaks between the two grids are below 0.03~eV.
In the pristine material, we observe the lowest transition at 2.45~eV.
This transition is dark and corresponds to the lowest indirect exciton.
It would not be captured in the unit cell, unless momentum transfers are considered as in Ref.~\citenum{palummo2020optical}. 
However, in the supercell used in this study, the X and L points are folded onto the $\Gamma$ point and are directly accessible by TD-DFT.
From the difference between this dark transition and the fundamental band gap in the polymorphous supercell of 2.83~eV we estimate the binding energy of the lowest indirect free exciton to be 0.39~eV.
This is slightly lower than the value of 0.48~eV, reported by Palummo~\textit{et al.}~\cite{palummo2020optical}.
The first bright transition, corresponding to the lowest direct free exciton is found at 2.96~eV, while the lowest direct independent-particle transition equals 3.57~eV in the same setup.
This implies the binding energy of the direct exciton of 0.61~eV. Within the unit cell, we find the binding energy to amount to 0.55~eV (see the related discussion in SI). We note that this value is significantly higher than previously reported ones (0.34~eV in Ref.~\citenum{palummo2020optical} and 0.17~eV in Ref.~\citenum{biega2021chemically}), which can be assigned to the higher fundamental band gap of \ce{Cs2AgBiBr6} in the present setup.
In their study, Biega~\textit{et al.}~\cite{biega2021chemically} demonstrated a linear relationship between the exciton binding energy and the size of the direct band gap, which would explain the higher value found here. In the SI we also demonstrate that by reducing the amount of exact exchange in the PBE0 functional, the exciton binding energy is reduced and in good agreement with previous studies in which the band gaps were also found to be smaller.

We now analyze the transitions in the TD-DFT calculations for the STE geometry.
We note that the calculation corresponds to the excitation from the singlet ground state.
However, since absorption and emission are inverse processes, the calculated energy transitions can be expected to be directly comparable to measured photoluminescence spectra.
The calculated transitions of the STE are given in Figure~\ref{fig:fig2}b).
While TD-DFT gives all transitions, we only plot those below 2~eV, as they correspond to the energy range in which photoluminescence is measured.
We compare the results with the experimental PL spectrum from Ref.~\citenum{wright2021ultrafast}.
The measurement was done at 4~K, hence the phonon broadening can be neglected in the comparison.
While the simulated spectrum is at lower energies than the experimental one, the difference between the midpoints of the spectra is only about 0.2~eV, representing fairly good agreement. 
We also note that the broadness of the experimental peak at low temperatures can be explained by the distribution of transitions which contribute to it. Previous works also suggested the possibility that the experimental emission is due to the indirect band gap~\cite{slavney2016bismuth, schade2018structural, palummo2020optical} or sub-gap states~\cite{hoye2018fundamental} such as color-centers~\cite{zelewski2019revealing, wright2021ultrafast}.
We rule out the emission from the indirect band gap based on two arguments. First, we find this transition at significantly higher energy, 2.49~eV. Second, this type of recombination requires an additional momentum change to occur. As for defects, to contribute to the emission, they would need to be deep and trap charges more strongly than the STE or small polarons. In that case, they would also lead to transitions at even lower energies than what we report here.
In Figure~\ref{fig:fig2}c) we show the band composition of the calculated transitions.
The lowest transition at 1.42~eV has a very low intensity and corresponds to a spin-forbidden triplet-singlet transition between the localized electron and hole states from within the STE.
Then, at 1.68~eV there is a bright transition from the singlet state of the STE.
This implies a separation of 0.26~eV between the lowest triplet and higher singlet states of the STE in \ce{Cs2AgBiBr6}.
At higher energies, between 1.74--1.95~eV, there is a distribution of transitions which all involve the localized electron state and both localized and delocalized hole states.
This range of energies can be used as an estimate of PL stemming from the electron polaron, if the STE dissociates.
In this energy range, we do not observe any transitions between the localized hole and delocalized electrons, because the hole in the STE has a more shallow level than the electron and this type of emission would mostly overlap with the absorption onset close to 3~eV. 

\begin{figure}[t]
    \centering
    \includegraphics[width=0.75\linewidth]{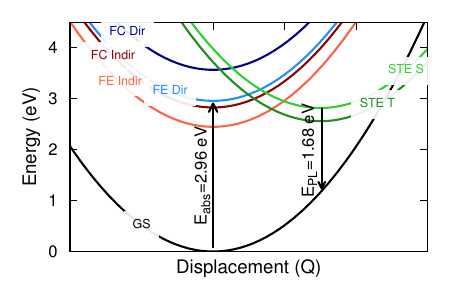}
    \caption{
    Configuration coordinate diagram of different electron-hole pairs in \ce{Cs2AgBiBr6}, including free carriers (FC), free excitons (FE) and self-trapped excitons (STE).
    `Indir' and `Dir' refer to indirect and direct transitions, and `S' and `T' to singlet and triplet states.
    }
    \label{fig:fig3}
\end{figure}

Using the energy transitions from the TD-DFT calculations, we finally construct the configuration coordinate diagram.
It allows us to compare the energies of direct, indirect and self-trapped excitons.
The diagram, given in Figure~\ref{fig:fig3} is based on TD-PBE0(28\%) calculations in the polymorphous cubic supercell. The energies at the ground-state and STE geometries are explicitly calculated and connected by parabolic curves to schematically represent the energy dependence on displacement.
The diagram reveals a complex landscape of electron-hole excitations in the material and can be used to analyze the dynamics of the excited charges.
First, direct free excitons are created.
These free excitons can then be trapped, in the form of the singlet STE.
This STE can then undergo four processes.
One, it can recombine leading to a bright emission that can be measured in photoluminescence.
Two, it can lower its energy and turn into a triplet form of STE.
Three, it can detrap and become an indirect free exciton, which has an energy very close to that of the triplet STE (within 0.1~eV in the current computational setup).
Finally, it could also dissociate leaving behind a localized electron and possibly a delocalized hole \cite{baskurt2023charge}. We note that this diagram provides a more comprehensive picture of the energetics involved compared to Ref.~\citenum{baskurt2023charge}, which only compared the energy of the triplet STE with the lowest indirect free-carrier transition.

In conclusion, we have studied different types of excitons in \ce{Cs2AgBiBr6} halide double perovskites.
We first assessed the performance of two non-empirical hybrid functionals, PBE0(28\%) and DDH in the description of the optical properties of the pristine material.
We have shown that both of them predict band gaps higher than what was previously reported in the literature and give absorption spectra in very good agreement with experiment.
We then used the TD-PBE0(28\%) technique to study excitations in the polymorphous cubic supercells corresponding to the ground-state structure and the self-trapped exciton.
This allowed us to show that the emission from the STE is in good agreement with experimental PL spectra.
Based on a configuration coordinate diagram, we finally revealed a complex landscape of electron-hole pairs in the materials with direct, indirect and self-trapped excitons having comparable energies.

\section*{Acknowledgments}
This work was supported by the Swedish Research Council (grant numbers 2019-03993, and 2020-04935), the Swedish Strategic Research Foundation through a Future Research Leader program (FFL21-0129) and the Wallenberg Academy Fellow program.
The computations were enabled by resources provided by the National Academic Infrastructure for Supercomputing in Sweden (NAISS) at C3SE, NSC, and PDC partially funded by the Swedish Research Council through grant agreements no. 2022-06725 and no. 2018-05973.

\end{document}